# PHEMENOLOGICAL MODELING OF ECLIPSING BINARY STARS

Andronov, Ivan L. [1]; Tkachenko, Mariia G. [1] and Chinarova, Lidia L.[2]

1) Department "High and Applied Mathematics", Odessa National Maritime University, Mechnikova, 34, 65029 Odessa, Ukraine, tt_ari@ukr.net, masha.vodn@yandex.ua

2) Astronomical Observatory, Odessa National University, Marazlievskaia, 1V, 65014 Odessa, Ukraine lidia_chinarova@mail.ru

**Abstract:** We review the method NAV ("New Algol Variable") first introduced in 2012Ap.....55..536A which uses the locally-dependent shapes of eclipses in an addition to the trigonometric polynomial of the second order (which typically describes the "out – of – eclipse" part of the light curve with effects of reflection, ellipticity and O'Connell). Eclipsing binary stars are believed to show distinct eclipses only if belonging to the EA (Algol) type. With a decreasing eclipse width, the statistically optimal value of the trigonometric polynomial *s* (2003ASPC..292..391A) drastically increases from ~2 for elliptic (EL) variables without eclipses, ~6-8 for EW and up to ~30-50 for some EA with narrow eclipses. In this case of large number of parameters, the smoothing curve becomes very noisy and apparent waves (the Gibbs phenomenon) may be seen. The NAV set of the parameters may be used for classification in the GCVS, VSX and similar catalogs. The maximal number of parameters is *m*=12, which corresponds to *s*=5, if correcting both the period and the initial epoch. We have applied the method to few stars, also in a case of multi-color photometry (2015JASS...32..127A), when it is possible to use the phenomenological parameters from the NAV fit to estimate physical parameters using statistical dependencies. For the one – color observations, one may estimate the ratio of the surface brightnesses of the components. We compiled a catalog of phenomenological characteristics based on published observations. We conclude that the NAV approximation is better than the TP one even for the case of EW – type stars with much wider eclipses. It may also be used to determine timings (see 2005ASPC..335...37A for a review of methods) or to determine parameters in the case of variable period, using a complete light curve modeling the phase variations. The method is illustrated on 2MASS J11080447-6143290 (EA-type), USNO-B1.0 1265-0306001 and USNO-B1.0 1266-0313413 (EW-type) and compared to various other methods from the literature.

## 1 Introduction

Numerous discoveries of the variable stars and the variety of their types of variability argue for a necessity of adequate mathematical methods, which will allow the best modeling. Of course, the best is to determine physical parameters of the stars (masses, radii, temperatures, magnetic field etc.). However, the majority of discoveries are being done using photometrical observations only (sometimes even with one or even no a filter). In this case, one may determine only so-called „phenomenological" parameters needed for determine its classification adopted in the „General Catalogue of Variable Stars" (GCVS, Samus', 2015),

„Variable Star Index" (VSX[1]) – co-ordinates, precession, maximum and minimum brightness, type of variability, and, if periodic, the period $P$ and the initial epoch $T_0$. An important parameter is $D$ (the full width of the minimum in eclipsing binaries in per cent of the period) or, alternately, $M - m$ (phase difference between the „Maxima" ($M$) and minima ($m$) for periodic pulsating variables, also in per cent of the period).

In the GCVS classicafication, the Algol-type stars (EA) have the property that „it is possible to specify, for their light curves, the moments of the beginning and end of the eclipses", contrary to other types – EB and EW, in the „light curves for which it is impossible to specify the exact times of onset and end of eclipses".

In this paper, we discuss the algorithm NAV („New Algol Variable") introduced by Andronov (2010, 2012) and illustrate it by an application to newly discovered variables – an EA-type system 2MASS J11080447-6143290 (Nicholson 2009), as the "main" star, and EW-type systems USNO-B1.0 1265-0306001 and USNO-B1.0 1266-0313413 (Hambsch 2007). More detailed description of the method and its application to other stars was presented by Andronov, Tkachenko & Chinarova (2016).

## 2 Methods of the analysis

### 2.1 Trigonometrical polynomials: asymmetric vs. asymmetric

We use a set of complementary methods. The periodogram analysis is carrried out using the trigonometric polynomial (TP) approximation of order $s$ of the light curve

$$x(t) = C_1 + \sum_{j=1}^{s} (C_{2j} \cos(2\pi f j(t - \bar{t})) + C_{2j+1} \sin(2\pi f j(t - \bar{t}))), \qquad (1)$$

where $\bar{t}$ is the sample mean of times of the observations, $f = 1/P$ - frequency, and $C_\alpha$ are the coefficients computed the method of the least squares (MLS). The test-function $S(f) = \sigma_C^2 / \sigma_O^2$, where $\sigma_C^2$, $\sigma_O^2$ are variances of the „calculated" and „observed" values at the times of the observations. The function was described by Andronov (1994, 2003) and has a sense of the square of the correlation coefficient between the observational and computed values (sometimes referred as $r^2$). The MLS is statistically justified, contrary to simplified methods of „Fourier Transform" (FT), which use formulae derived for a restricted case of equidistant data only. From mathematical point of view, this means that some or many elements of the normal equations are set to zero instead of being computed and taken into account (e.g. Deeming, 1975, Lomb, 1976, Scargle, 1982). Mikulášek (2007) noted this as a „frequent syndrome of variable stars observers, which could be named *Matrixphobia*".

---

In our program, the coefficients are determined using correct formulae. Moreover, our program „Multi-Column View" (MCV), Andronov & Baklanov (2004) allows to make periodogram analysis with taking into account the possible algebraic polynomial trend of needed degree, contrary to the method of „prewhitening", a kind of methods with a „*Matrixphobia*".

In Fig. 1, the periodograms for some degrees of the trigonometrical polynomial fit are shown. One may see that there is no peak at the true period for $s = 1$, but occurs at $s \geq 2$.. This is explained by two nearly equal eclipses seen in Fig. 2, so formally the „better" photometric period is twice smaller than the true one. With an increasing $s$, the height and number of peaks increase. It should be noted, that the peak at the true period remains lower than at the half-period even for a relatively large degree of the trigonometrical polynomial $s = 8$. This is explained by a very narrow eclipse (what is indeed typical for the EA – type systems), because much larger values of are needed for better approximation.

Another problem is to determine the statistically optimal degree of the trigonometrical polynomial $s$. Andronov (1994, 2003) and Andronov & Marsakova (2006) discussed criteria based on:

1. the Fischer statistics with setting the limiting value of the „False Alarm Probability" (FAP);

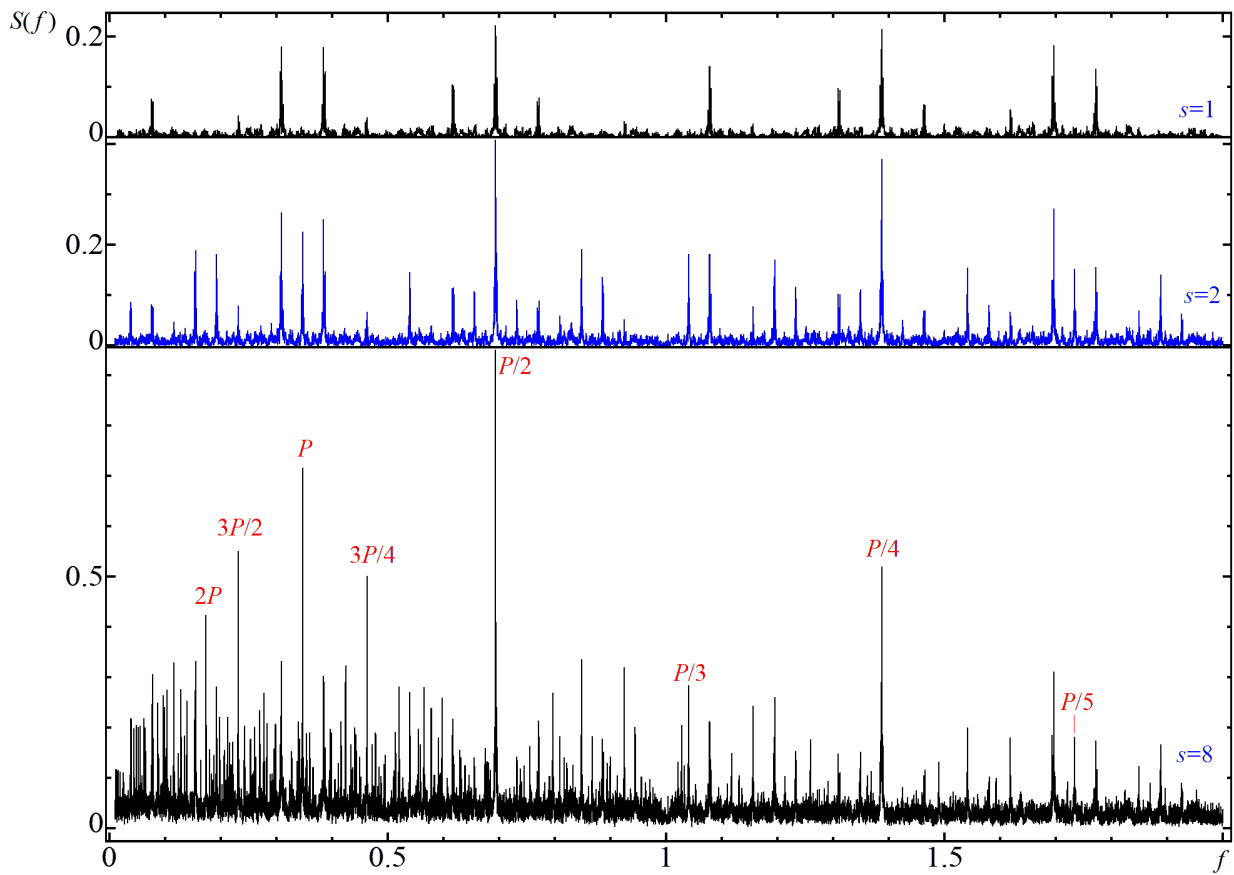

Figure 1: Periodograms $S(f)$ using trigonometric polynomial fits of orders $s = 1, 2, 8$. The positions of peaks related to the main period are marked.

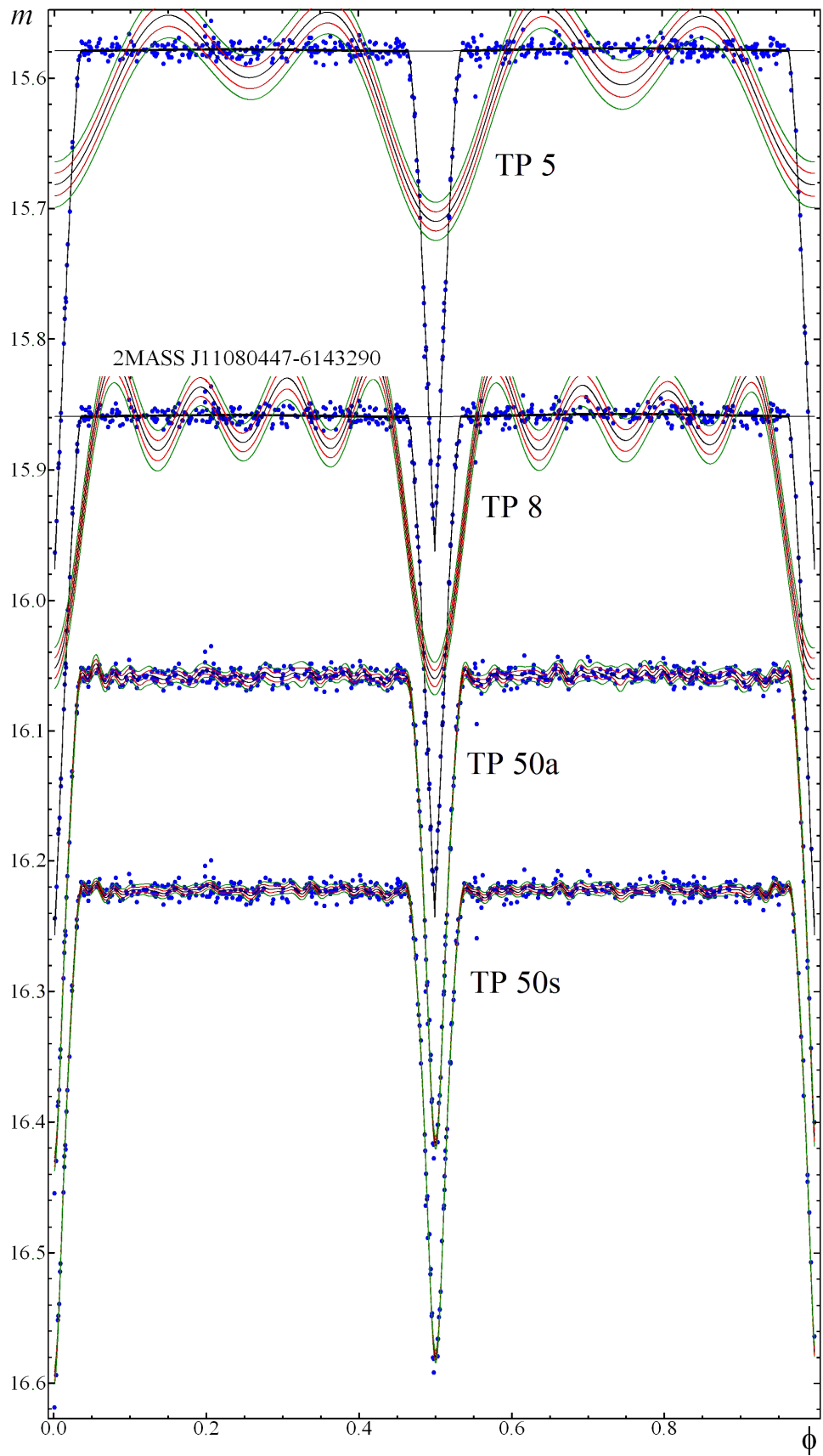

Figure 2: Trigonometric polynomial approximations of orders $s = 5, 8, 50$ („a" − asymmetric and „s" − symmetric). The „$\pm 1\sigma$" and „$\pm 2\sigma$" error corridors are shown in red and green.

2.  the minimization of the error estimate (with modifications – r.m.s. error of the smoothing function at all observations $\sigma[x_C]$; at some specific time or phase $\sigma[x_C(t_0)]$; or of the time of the extremum $\sigma[t_e]$);

3.  the maximum „signal-to-noise" ratio $\sigma_C / \sigma[x_C]$, i.e. the ratio of the r.m.s. deviation of observations from the sample mean to the r.m.s. error of the approximation.

In Fig. 2, the TP approximations are shown for different values of the parameter $s$, which argue for larger number of $s$. also we show the „large order" fits in Fig. 3 is shown the dependence of the coefficients of the „symmetric" trigonometric polynomial fit

$$x(t) = C_1 + \sum_{j=1}^{s} C_{j+1} \cos(2\pi f j(t - T_0)), \tag{2}$$

where $T_0$ – is the „initial epoch", which, in this case, is the point of symmetry of the calculated light curve. Generally, the asymmetric fit is needed to check the presence of the asymmetry (the O'Connell (1951) effect). For this star, the asymmetry is not statistically significant, thus we do not show the coefficients. Also the even coefficients $C_{2k+1}$ do not exceed the „$\pm 3\sigma$" corridor and formally are not statistically significant. They describe the difference in the shape of the minima, so they are small for this star with similar minima. This is not the case for the systems with significantly different minima, e.g. 2MASS J11080308-6145589 (Andronov, Tkachenko & Chinarova, 2016). However, the cumulative effect of these coefficients is present, and there is a small difference in depth between the two minima, as is seen in Fig. 2, and will be discussed below.

Very interesting is a dependence of the even coefficients $C_j$, $j = 2k$ on $j$. It resembles a one-sided gaussian, and is well approximated as

$$C_j = A \exp(-Bj^2) \tag{3}$$

with the parameters $A = 0.0459, B = 0.00345$ obtained using non-linear MLS. This differs for a nearly linear decrease seen for another EA – type star 2MASS J11080308-6145589 at Fig. 4 of Andronov, Tkachenko & Chinarova (2016).

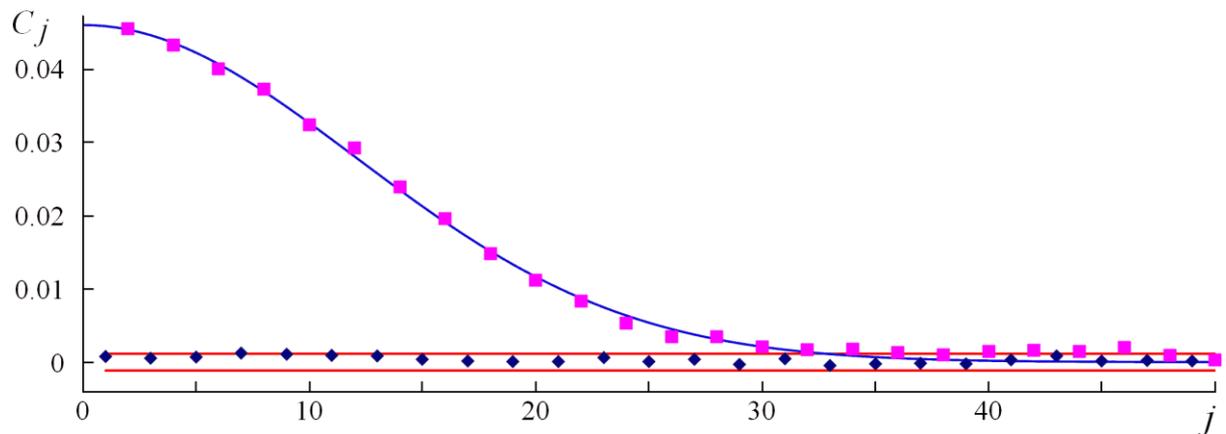

Figure 3: The coefficients of the „symmetric" TP approximation: the even ones $C_{2k}$ are shown by squares and the odd ones $C_{2k+1}$ – by rombs. The red horizontal lines show the „$\pm 3\sigma$" corridor. The blue line is a gaussian – type approximation.

Using the „$\pm 3\sigma$" criterion, the statistically optimal degree is $s = 46$. If avoiding the coefficients within the wider „$\pm 8.9\sigma$" corridor, one may suggest a significantly smaller value $s = 28$. However, even using a larger value $s = 50$, the approximations do not fit well in depth both minima (Fig. 2). Similarly to Fig. 8 of Andronov et al. (2016), the brightness at minima (both primary and secondary) of the approximation decreases with $j$, even if the coefficients are statistically not significant.

## 2.2  Combined „constant – line" approximation

For a comparison, in Fig. 4 we show the „sampling" approximation of the phase light curve using 50 sub-intervals, as typically used as the input data for the programs of physical modeling (e.g. Zoła, Kolonko. &  Szczech, 1997, Zoła, S., Gazeas, K., Kreiner, J.M. et al., 2010). Contrary to an usual method of computing sample mean values in each sub-interval, and creating a set of weighted „observations" ($\bar{t}, \bar{x}, \sigma[\bar{x}]$), in the program MCV (Andronov & Baklanov, 2004), we realized a combined method: the data in the current sub-interval are approximated by a constant and by a line (algebraic polynomials of the degrees $p = 0$ and $p = 1$, respectively). Then the approximation is chosen, for which the error estimate of the smoothing function at the mean moment of observations is smaller.

This definitely has an advantage at the intervals of drastic changes (ascending and descending branches). At the out-of-eclipse phases, the line becomes „better" for a few times, when the number of points in the interval is too small ($2 - 3$), so statistical errors have bad accuracy themselves. The resulting table of the observations remains the same for the line and constant, but the error estimates at the ascending and descending branches become more realistic, because systematic deviation of the curve from a constant overestimates the stastistical error, and so underestimates the weight of the mean point.

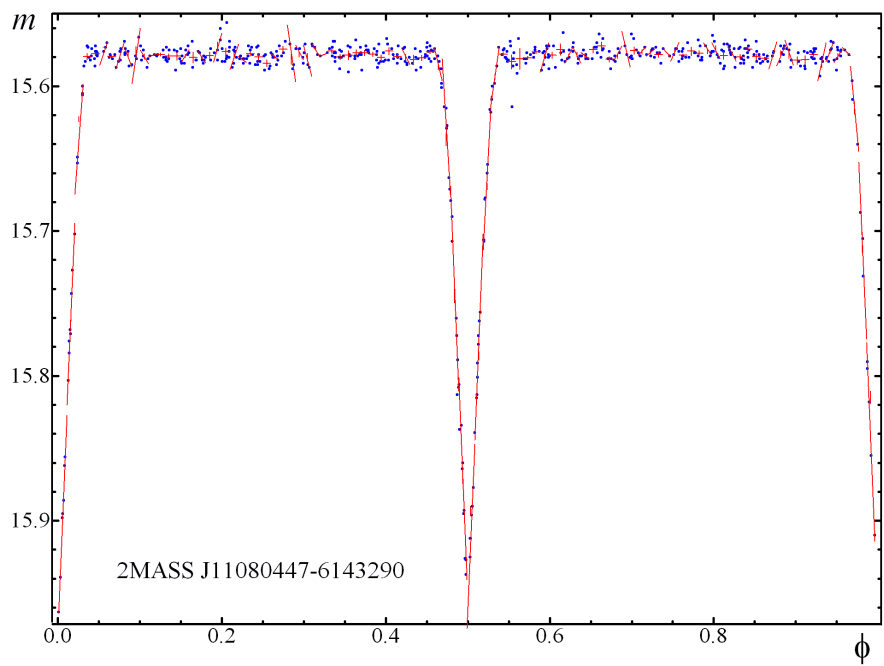

Figure 4.
Approximation of the phase light curve (blue points) by a constant / line in 50 phase intervals (red lines).

Because the lines from individual sub-intervals do not form a joint continuous function, the approximation may be described as the „polynomial spline of degree 1 and defect 2". An application of splines to variable stars was discussed by Andronov (1987). However, in future analysis, only a discrete set of such averaged observations is needed, thus this discontinuity of the smoothing curve is replaces by a continuous light curve obtained from phenomenological or physical modeling.

Similar approximation was used for fast periodogram analysis by Marraco & Muzzio (1980). They have used only linear approximations, but in MCV the statistically better (among „constant" or „line") approximation is chosen.

Obviously, the mean points computed in this manner, can not be outside of the interval of observational values. However, the extrapolation in the sub-intervals close to the minimum can, and we see sharp „triangle – shaped" minima, contrary to the more smooth trigonometric polynomials.

## 2.3   Comparison of the  approximations of the eclipses

For the determination of the moments of eclipses, which is the major activity of studies at small telescopes, the most popular method in the computer era is the method of Kwee and van der Woerden (1956), which is implemented in the commercial software PERANSO[2] . This method makes a „saw-tooth"-like (spline of power 1 and defekt 1) interpolation of the observations in the user-determined interval (near the extremum), and then a search of the „symmetry point" is determined with a discontinuous derivative, what may significantly underestimate the statistical error. So Mikulášek, Chrastina, Liška et al. (2014) recommend to „retire" this method.

Sometimes may occur an error due to a „division by a zero", if the observations contain points with equal arguments (e.g. observations at the same time from different observers) or phases (e.g. determination of the initial epoch for the phase light curve containing observations from different cycles of variability). In our realization of the method in the program „Observation Obscurer" (first version introduced by Andronov (2001)), the points with equal arguments are averaged.

Despite formally one may determine an interpolated value of the extremum, the error estimate of the interpolated value is typically is not defined.

Other methods are based on the least squares (e.g. Anderson 1958, Andronov 1994, 2003, Mikulášek, 2007, 2015, Chrastina, Mikulášek, & Zejda, 2014). Kurochkin (1963), Nikonov (1971) and many further authors recommend to use for the approximation of the observation in the interval around trial extremum by an algebraic polynomial of degree  $p$

$$x(t) = C_1 + \sum_{j=1}^{p} C_{j+1}(t - T_p)^j,$$  (4)

where $T_p$  is some time close to (or inside) the interval near the extremum.

---

Then the moment of extremum may be found by solving the equation $x'(t_e) = 0$, e.g. using the iterations $t_e := t_e - x'(t_e)/x''(t_e)$ with an error estimate $\sigma[t_e] = \sigma[x'(t_e)]/|x''(t_e)|$ (e.g. Andronov, 1994, 2005, Mikulášek, 2007, 2015). Contrary to the software PERANSO, which asks the user for the degree of the polynomial $p$, we prefer to use the statistically optimal value, which corresponds to the best accuracy of the timing $\sigma[t_e]$. Using this algorithm, Chinarova & Andronov (2000) determined characteristics of 6509 extrema, which are the part of their „Catalogue of Main Characteristics of Pulsations of 173 Semi-Regular Stars".

Another method of „asymptotic parabola" fit was developed by Marsakova & Andronov (1996), where the smoothing function is a spline with different degree (1;2;1), or two lines connected with a parabola in such a way that the function and its first derivatives are continuous. This method is effective for pulsating variables with asymmetric extrema, but needs determination of the interval near each extremum, which has nearly linear parts of the ascending and descending branches. Recently Andrych, Andronov, Chinarova et al. (2015) realized both these methods in the VBA program. A comparison of the algebraic polynomial vs. „asymptotic parabola" may be seen in Fig. 5.

An extension of the method of algebraic polynomials is again a spline with a special shape $H(z) = H(-z)$, which is monotonically decreasing from $H(0) = 1$ to $H(1) = 0$ and $H(z) = 0$ for $|z| \geq 1$. Here $z = (\phi - \phi_0)/\Delta\phi$, where $\phi_0$ is the phase of symmetry of the minimum (typically 0 and 0.5 for the primary and secondary minimum, respectively, but may be corrected, if shifted), and $\Delta\phi$ is a half-duration of the eclipse.

The simplest form of such function may be $H(z) = 1 - |Z|^{\beta}$. In Fig. 6, the approximations are shown for $\beta = 1$ (triangle), $\beta = 2$ (parabola) and also symmetric polynomial of degrees 4 ($H(z) = 1 - (1 - D_4)z^2 - D_4 z^4$) and 6 :

$$H(z) = 1 - (1 - D_4 - D_6)z^2 - D_4 z^4 - D_6 z^6 \tag{5}$$

Here we used an „out-of-eclipse" part of the curve, which is approximated (for this star with flat maxima) by a constant, so $x(\phi) = C_1 + C_2 H((\phi - C_3)/C_4)$.

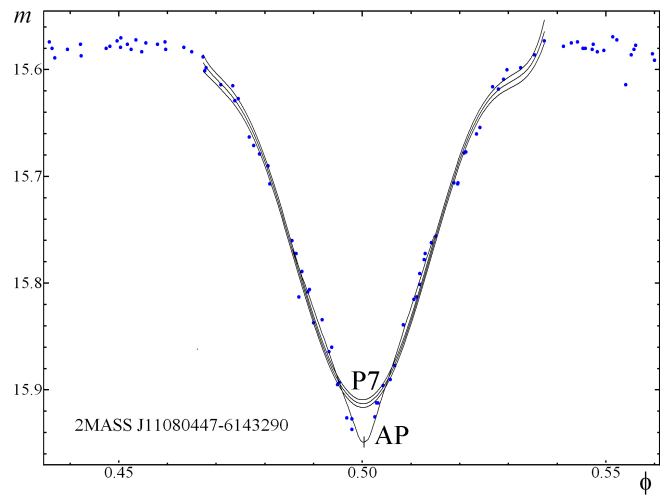

Figure 5. Approximation of the part of the phase light curve near the secondary minimum (blue points) by the polynomial with 7 parameters (6-th order) „P7", for which the „$\pm 1\sigma$" error corridor is shown, and the „asymptotic parabola" („AP") fit.

Moreover, the coefficients in Eq. (5) may be renamed as $C_5 = D_4, C_6 = D_6$. Even the visual comparison shows that the best approximations are „S1" and „S6". Next two fits are based on the gaussian: $H(z) = \exp(-z^2/2)$ and its highly improved version listed in Eq.(14) of Mikulášek (2015):

$$H(z) = (1 + C_6 z^2 + C_7 z^4)(1 - (1 - \exp(1 - \cosh(z)))^{C_5}) \qquad (6)$$

Generally, one may expect that, with an increasing number of the parameters, the quality of the fit will increase. And this was exelently illustrated by Mikulášek (2015) for a case of short ascending/descending branches and a „shallow parabola-like" minimum during a planet transit in the system HD209458b.

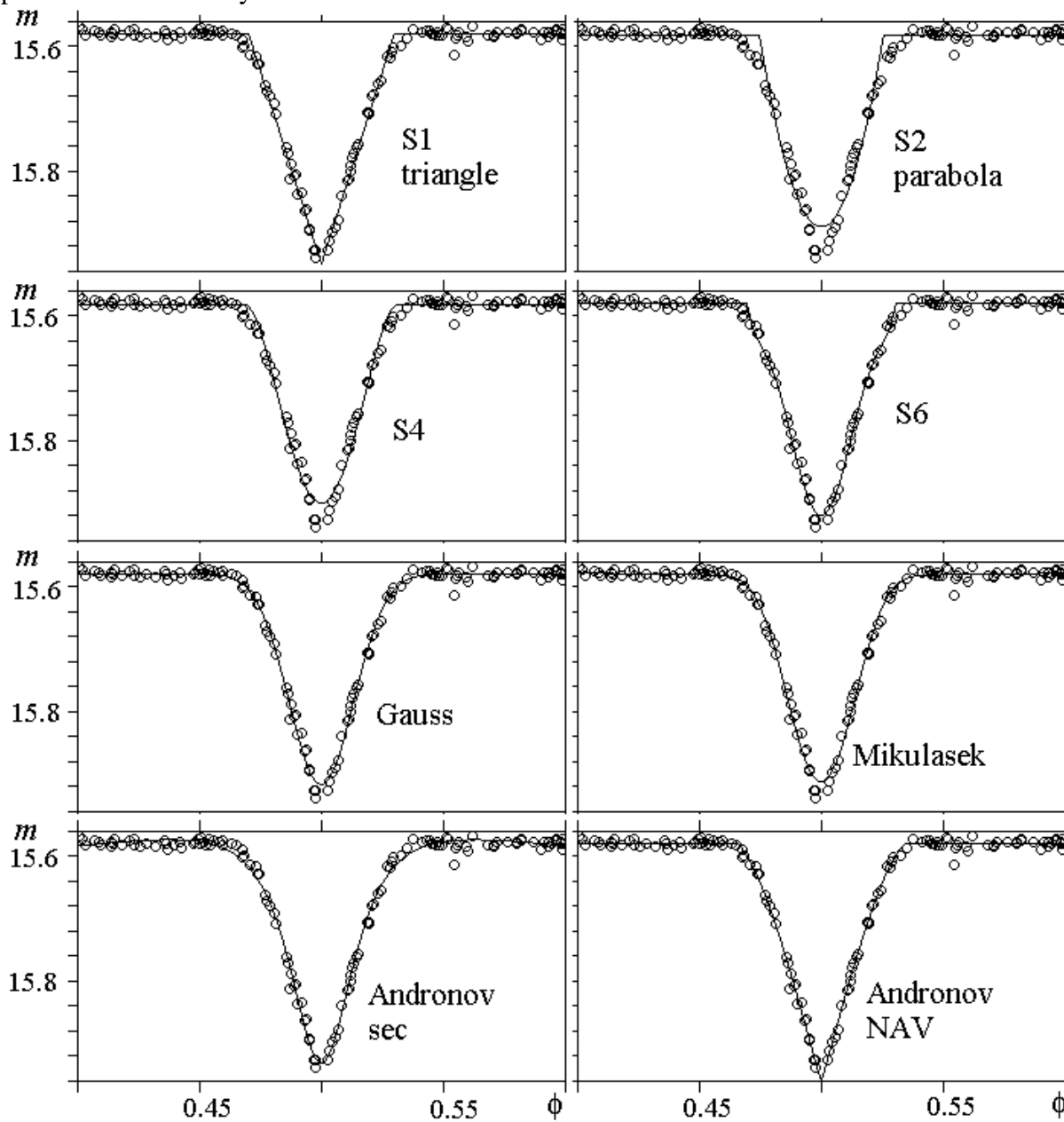

Figure 6: The approximations of the vicinities of the secondary minimum using various approximations described in the text.

However, for our example of the EA – type star with no „plateau" at the bottom of the eclipse, the function (6) describes the light curve (particularly, the depth of the minimum) even worse than the gaussian. The approximations „S1" and „S6" are worse for a complete light curve, but better in depth of the minimum.

Another extension of the hyperbolic function was proposed by Andronov (2005) for the case of the asymmetric maxima (pulsations or outbursts), assuming asymptotically exponential rise and decay:

$$H(z) = \frac{1}{\exp(-z) + \exp(+C_5 z)} \tag{7}$$

In the case of symmetric minima (assumed for eclipsing binaries without O'Connell effect), this function may be simplified to $H(z) = 1/(2\cosh(z))$, or, better, to scale by a factor of 2: $H(z) = 1/\cosh(z) = \mathrm{sech}(z)$. The corresponding fit is worse near the beginning/end of the eclipse, but better at the center of the eclipse. One may try other modifications with additional parameters (in an order of decreasing importance) like

$$H(z) = (1 + C_7 z^2 + C_8 z^4)\mathrm{sech}^{C_5}(|z| + C_6 z^2). \tag{8}$$

Both the gaussian and the functions (6), (7), (8), despite good approximating properties, do not allow to determine the width of the minimum (which is one of the main parameters „$D$" needed to register the star in the „General Catalogue of Variable Stars" (Samus', 2015)), as the corresponding functions $H(z)$ formally reach zero only asymptotically, but not at the finite values of $|z| = 1$. To solve this problem, Andronov (2010, 2012) proposed to use a locally defined function, and, among few, had chosen

$$H(z) = (1 - |z|^{C_5})^{3/2}. \tag{9}$$

This function has a correct asymptotic behaviour at $|z| \to 1$, and was found to be effective for dosens of eclipsing binaries already studied. It seems also to be the best in Fig. 6. This phenomenological approximation was called „NAV" („New Algol Variable").

The initial algorithm for a fixed phase curve (i.e. the period and the initial epoch determined using other methods, e.g. the trigonometrical polynomial fit of statistically optimal degree) was improved in such a way, that the period and the epoch are determined as additional „non-linear" parameters. Obviously, one may add terms taking into account any model of the period variations similar to other fits (cf. Andronov, 1987, Mikulášek, 2015) .

The fits and raw graphs shown in Fig. 6 were obtained using the trial version of the software WinCurveFit v. 1.1.2 (Kevin Raner Software), which realized determination of the coefficients of the non-linear models. Unfortunately, the program works unstable, and the solution often does not converge even with good initial point. Other fits were computed using MCV (Andronov & Baklanov, 2004) and other own programs. In the Table 1, we list some characteristics of some fits for a comparison.

As the quality of the fit we use the value of $r^2$ (best, if equal to unity), but, for the quality of the fit near the mid-eclipse, we use the value at the minimum and redefine $C_1 = x(\phi_0)$. In this case, we introduce the function $G(z) = 1 - H(z)$, so $x(\phi) = C_1 - C_2 G((\phi - C_4)/C_3)$ .

Table 1: Characteristics of the fits of observations near the secondary minimum. The error estimates are in parenthesis in units of the last decimal digit.

| $G(z)$ | $C_1 = x(\phi_0)$ | $C_2$ | $C_3 = \Delta\phi$ | $C_4 = \phi_0$ | $C_5$ | $C_6$ | $\sigma_0$ |
|---|---|---|---|---|---|---|---|
| NAV: $1-(1-|z|^{C_5})^{3/2}$ | 15.9586(50) | 0.3794(52) | 0.03489(41) | 0.499908(84) | 1.083(30) | | 0.99581 |
| sech2: $(1-\mathrm{sech}(z))(1+C_5z^2)$ | 15.9312(32) | 0.3614(35) | 0.01135(19) | 0.499970(116) | -0.00073(17) | | 0.99253 |
| sech1: $(1-\mathrm{sech}(z))$ | 15.9346(34) | 0.3609(34) | 0.01082(15) | 0.499979(124) | | | 0.99121 |
| Sech3: $(1-\mathrm{sech}(z))^{C_5}$ | 15.9175(42) | 0.3420(46) | 0.00936(34) | 0.499981(118) | 1.415(115) | | 0.99229 |
| Kwee &van Woerden (1956) | | | | 0.499905(167) | | | |
| S1: $|z|$ | 15.9483(25) | 0.3684(26) | 0.02985(23) | 0.499932(102) | | | 0.99381 |
| S2: $z^2$ | 15.8884(40) | 0.3055(43) | 0.02534(31) | 0.499965(208) | | | 0.97538 |
| S4: $(1-C_5)z^2 + C_5z^4$ | 15.8994(31) | 0.3185(34) | 0.03148(85) | 0.499887(157) | -1.00(11) | | 0.98748 |
| S6: $(1-C_5-C_6)z^2 + C_5z^4 + C_6z^6$ | 15.9192(28) | 0.3393(30) | 0.0304(33) | 0.499956(105) | -3.51(21) | 1.73(14) | 0.99382 |
| Gauss: $(1-\exp(-z^2/2))$ | 15.9166(24) | 0.3379(25) | 0.01321(13) | 0.499932(106) | | | 0.99412 |
| Mikulášek: $(1-\exp(1-\cosh(z)))^{C_5}$ | 15.9164(28) | 0.3377(31) | 0.1356(9746) | 0.499931(107) | 105(1512) | | 0.99411 |
| Mikulášek: $(1-\exp(1-\cosh(z)))^{C_5}(1+C_6z^2)$ | 15.9165(28) | 0.3372(40) | 0.1676(20966) | 0.499932(107) | 161(4038) | 0.010(236) | 0.99411 |
| $(1-\exp(1-\cosh(z)))^{C_5}(1+C_6z^2+C_7z^4)$ | 15.9159(28) | 0.3407(70) | 0.1251(10409) | 0.499930(107) | 88(1461) | -0.086(1506) | 0.99416 |
| Asymptotic parabola ($m = 5$) | 15.9489(38) | | | 0.500223(20) | | | |
| Polynomial (best, $m = 7$ parameters, 6th degree) | 15.9126(36) | | | 0.499996(203) | | | |
| Symmetric polynomial ($m = 10$, 18th degree) | 15.9385(46) | | | 0.499934(95) | | | |
| Trigonometric polynomial ($m = 93$, 46th degree) | 15.9244(21) | | | | | | |

Some fields are missing because the sets used for different methods are different: the trigonometrical polynomial fit is computed for all the data ($n = 560$); the asymptotic parabola – for the smallest interval near mid-eclipse (points 276..305 after sorting in phase), where the ascending/descending branches were nearly linear; the remaining approximations were done for the points 264..319. The observations were sorted according to phases computed for the ephemeris (Table 2) obtained for the NAV fit for all data. The best accuracy (smallest $\sigma[\phi_0]$) is estimated for the asymptotic parabola fit, the second one in this reyting is the NAV fit, which is preferable because of using the complete light curve and thus all phenomenological parameters may be determined. Previous studies (Andronov, 2012, Andronov, Tkachenko and Chinarova, 2016) had shown that the NAV fit typically has much smaller r.m.s. statistical error than the trigonometric polynomial, even if the accuracy of the curve is nearly the worst near the phases of mid-eclipses due to sharper shape than that of the trigonometrical polynomial.

A comparison of the exponent-based fits shows that the complification of the basic functions (the gaussian, the hyperbolic secant) makes minor improvements to the quality of the fit, but the error estimates of the additional parameters typically exceed these parameters by a factor of many times.

## 3   Discussion of the NAV approximation of the complete curve

As was shown above, the basic function („special shape") for the eclipse is $H(z;\beta) = (1 - |z|^\beta)^{3/2}$, $-1 \leq z \leq +1$, where $\beta = C_5$ is the parameter describing behaviour close to the mid-eclipse (0 – very narrow, 1 – triangular, 2 – parabolic, $\gg 2$ – flat).

The complete function includes a TP2 part (a trigonometrical polynomial of the second order), which approximates three effects: reflection, ellipticity and O'Connell and has 12 parameters, including two for the corrected initial epoch and the period (Andronov, Tkachenko & Chinarova, 2016):

$$x(\phi) = C_1 + C_2 \cos(2\pi(\phi - \phi_0)) + C_3 \sin(2\pi(\phi - \phi_0)) + C_4 \cos(4\pi(\phi - \phi_0)) + C_5 \sin(4\pi(\phi - \phi_0)) +$$
$$+ C_6 H((\phi - \phi_0)/C_8; C_9) + C_7 H((\phi - \phi_0 - 0.5)/C_8; C_{10}). \tag{10}$$

In previous works (Andronov, 2012), we used $\phi_0 = 0$, but, in this recent work, we added two parameters $(C_{11}, C_{12})$ to correct the initial epoch and the period. Of course, when needed, it is possible to add more parameters to describe the possible period changes.

Some of these and related parameters are listed in Table 2. The corresponding light curves are shown in Figure 7. It is clearly seen, that the eclipses are well pronounced not only for the EA star, but also for two EW-type variables.

This is also seen in 6 more stars analyzed by Tkachenko, Andronov & Chinarova (2015). The „zero amplitudes" of the „eclipse shapes" appear only in low-amplitude stars, where the variability is due to ellipsoidality only, without any eclipses. An additional advantage of such parametrization is also a possibility to estimate the degree of eclipse $Y = d_1 + d_2$ (equal to zero, if no eclipses, and unity, if both eclipses are total) and the ratio of the mean brightnesses at the maximum phase of primary and secondary eclipses $\gamma = d_1 / d_2$. If neglecting the limb darkening, this will be the ratio of the brightnesses of the limbs of two stars, so one may estimate physical parameters, as the „first guess" (="initial values") for the physical modeling. Details and references to this model, may be found in Tkachenko & Andronov (2013).

Table 2: Phenomenological characteristics of the eclipsing variable stars.

| Star | 2MASS J11080447-6143290 | USNO-B1.0 1265-0306001 | USNO-B1.0 1266-0313413 |
|---|---|---|---|
| Reference | Nicholson (2009) | Hambsch, 2007 | Hambsch, 2007 |
| $T_0$ | 2450941.22839(16) | 2454300.36951(28) | 2454312.87437(12) |
| $P$ | 2.88349387(38) | 0.57930071(41) | 0.3003726(34) |
| $\Delta\phi$ | 0.03480(27) | 0.11901(131) | 0.10902(120) |
| $Y = d_1 + d_2$ | 0.5978(43) | 0.2686(46) | 0.2548(36) |
| $\gamma = d_1 / d_2$ | 1.0288(125) | 1.5389(337) | 1.0971(203) |

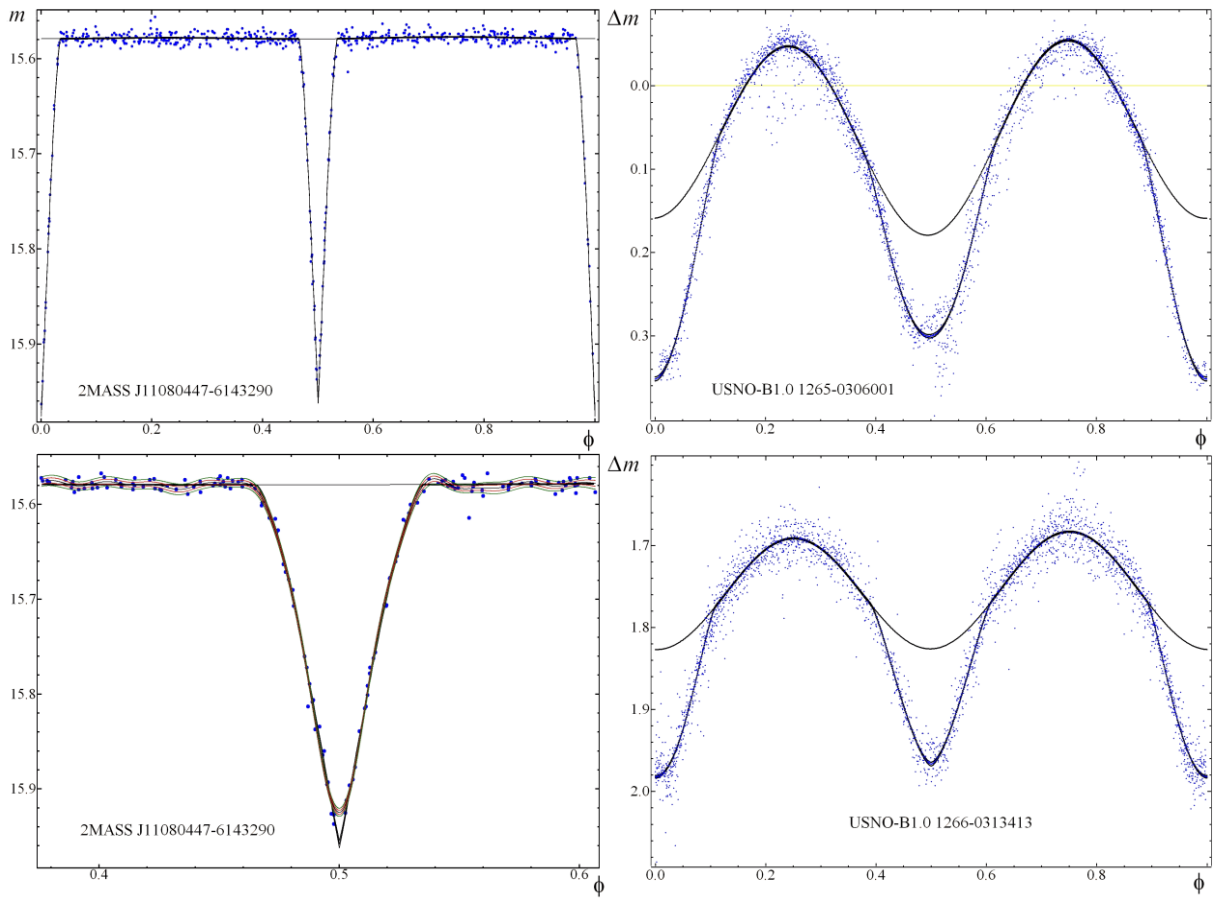

Figure 7. The NAV fits for the EA-type star (left) and two EW-type stars (right). At the bottom left figure, the vicinities of the secondary minimum are shown with the NAV and TP46 fits. The line above the fit at the phase interval of the eclipse is a TP2 part of the Eq. (10), which describes possible reflection, ellipticity and the O'Connell effects.

## 4    Multi-color observations

While searching the „Open European Journal on Variable Stars"[3] for multicolor observations of eclipsing binaries (3 − 4 filters), we have found only the papers by Juryšek & Hoňková (2012) and Devlen (2015). Such observations are important to determine color indexes and to estimate temperatures of the stars, what allows to make estimates of the physical parameters using phenomenological modeling. An example of such an analysis using the statistical „mass-radius-temperature-luminosity" relations was presented by Andronov, Kim, Kim et al. (2015) for the system 2MASS J18024395 + 4003309 = VSX J180243.9+400331.

## 5    Summary

Different methods for an approximation of the light curve either „global" (i.e. a complete light curve), or local (intervals near minima), are compared. The characteristics of variability of the three stars chosen for the analysis, show an efficiency of the NAV algorithm. Even if more complicated models may be comparable in accuracy, the smallest possible number of the NAV parameters and their cleas physical sense is the advantage. The corrected values of the period and the initial epoch are more accurate than that obtained by other methods, sometimes by a factor of many times. Thus the NAV algorithm is recommended to be used for phenomenological modeling of newly discovered or poorly studied stars, for which the physical modelling is impossible (in the sense of solution in the region of the parameter space instead of the single point) because of unknown temperatures and mass ratios.

In future, we plan to compile a catalogue of characteristics of a group of eclipsing binaries with published original observations − either that needed for the GCVS, or additional ones proposed while applying the method.

## References

Anderson, T.W. 1958, An Introduction to Multivariate Statistical Analysis, J.Wiley & Sons, 374p.

Andrych, K.D., Andronov, I.L, Chinarova, L.L., Marsakova, V.I. 2015, Odessa Astron. Publ., 28, 158, 2015OAP....28..158A

Andronov, I.L. 1987, Publ. Astron. Inst. Czechoslovak. 70, 161, 1987PAICz..70..161A

Andronov, I.L. 1994, Odessa Astron. Publ., 7, 49, 1994OAP.....7...49A

Andronov, I.L. 2001, Odessa Astron. Publ., 14, 255, 2001OAP....14..255A

Andronov, I.L. 2003, ASP Conf. Ser., 292, 391, 2003ASPC..292...391A

Andronov, I.L. 2005, ASP Conf. Ser., 335, 37, 2005ASPC..335...37A

Andronov, I.L. 2010, Int. Conf. KOLOS-2010 Abstract Booklet, 1, http://www.astrokarpaty.net/kolos2010abstractbook.pdf

Andronov, I.L. 2012a, Astrophysics, 55, 536, 2012Ap.....55..536A

Andronov, I.L. 2012b, Częstochowski Kalendarz Astronomiczny 2013, ed. B.Wszołek & A. Kuźmicz, 133, 2012arXiv1212.6707A , http://ptma.ajd.czest.pl

---

[3] http://var.astro.cz/oejv

Andronov, I.L. & Baklanov, A.V. 2004, Astronomy School Reports, 5, 264, 2004AstSR...5..264A , http://jrnl.nau.edu.ua/index.php/ASTRO/article/view/9343/11641

Andronov, I.L., Kim, Yonggi, Kim, Young-Hee, Yoon, Joh-Na, Chinarova, L.L.& Tkachenko, M.G. 2015, Journal of Astronomy and Space Science, 32, 127, 2015JASS...32..127A

Andronov, I.L. & Marsakova, V.I. 2006, Astrophysics, 49, 370, 2006Ap.....49..370A

Andronov, I.L.& Tkachenko, M.G. 2013, Odessa Astron. Publ., 26, 204, 2013OAP....26..204A

Andronov, I.L., Tkachenko, M.G. & Chinarova, L.L. 2016, Phys.J., 2, 140, 2015arXiv151000333A , http://files.aiscience.org/journal/article/pdf/70310096.pdf

Chinarova, L. L. & Andronov, I. L. 2000, Odessa Astron. Publ., 13, 116, 2000OAP....13..116C

Christina, M., Mikulášek, Z. & Zejda, M. 2014, Contributions of the Astronomical Observatory Skalnaté Pleso, 43, 422, 2014CoSka..43..422C

Deeming, T.J. 1975, Ap&SS, 36, 137, 1975Ap&SS..36..137D

Devlen, A. 2015, OEJV, 171, 1, 2015OEJV..171....1D

Hambsch, F.-J. 2007, OEJV, 67, 1, 2007OEJV...67....1H

Juryšek, J. & Hoňková, K. 2012, OEJV, 152, 1, 2012OEJV..152....1J

Kwee, K. K. & van Woerden, H. 1956, Bull. Astron. Inst. Netherlands, 12, 327, 1956BAN....12..327K

Lomb, N.R. 1976, Ap&SS, 39, 447, 1976Ap&SS..39..447L

Marraco, H. G. & Muzzio, J. C. 1980, PASP, 92, 700, 1980PASP...92..700M

Mikulášek, Z. 2007, OAP, 20, 138, 2007OAP....20..138M

Mikulášek, Z. 2015, A&A, 20, 584A, 8, 2015A&A...584A...8M

Mikulášek, Z., Christina, M., Liška, J., Zejda, M., Janík, J., Zhu, L.-Y. & Qian, S.-B., 2014CoSka..43..382M

Nicholson, M.P. 2009, OEJV, 102, 1, 2009OEJV..102....1N

O'Connell, D.J.K. 1951, Publ. Riverview Coll. Obs., 2, 85, 1951PRCO....2...85O

Tkachenko, M.G., Andronov, I.L. & Chinarova, L.L. 2015, Odessa Astron. Publ., 28, 181, 2015OAP....28..181T

Samus', N.N. (ed.) 2015, General Catalog of Variable Stars, http://www.sai.msu.su/gcvs/gcvs/ , 2009yCat....102025S

Scargle, J.D. 1982, ApJ, 263, 835, 1982ApJ...263..835S

Zoła, S., Kolonko, M. & Szczech, M. (1997). Analysis of a Photoelectric Light Curve of the W UMa-Type Binary ST Ind. A&A, 324, 1010. 1997A&A...324.1010Z

Zoła, S., Gazeas, K., Kreiner, J.M., Ogloza, W., Siwak, M., Koziel-Wierzbowska, D. & Winiarski, M. 2010, MNRAS, 408, 464, 2010MNRAS.408..464Z